# High degree of chaos synchronization in pairs of transverse modes with parity-symmetric polarizations in a thin-slice solid-state vector laser and application to polarimetric secure self-mixing metrology

**Kenju Otsuka**

*TS³L Research, 126-7 Yamaguchi, Tokorozawa, Saitama 359-1145, Japan*
*Corresponding author: Kenju.otsuka@gmail.com*

**A high degree of chaos synchronization among transverse mode pairs with parity-symmetric polarizations was demonstrated in a laser-diode pumped thin-slice c-cut Nd: GdVO₄ laser. The amplitude correlation coefficient greater than R > 0.994 resulted from a non-reversing mirror symmetric-polarization-dependent structural change in the modal patterns, where chaotic dynamics of the whole system consisting of qualitatively different dynamics, which depend on the polarization crossing angle, resembled that of a chaotic single mode laser. A self-organized sender-receiver type of chaos synchronization of a single pair of modes among an infinite number of parity-symmetric polarizations and the associated chaotic camouflaging as well as extracting experiment of a self-mixing solid-state laser Doppler velocimetry signal were demonstrated. © 2020 Optical Society of America**

http://dx.doi.org/10.1364/OL.99.099999

A variety of chaos synchronization phenomena have been reported in solid-state lasers subjected to a pump or loss modulation at a frequency close to the relaxation oscillation frequency below few MHz. The synchronization is established in the form of spatially coupled phase-locked arrays [1] or in the form of unidirectional injection locking [2-6]. As well, synchronization of chaos in semiconductor laser diodes (LDs) subjected to delayed feedback [7] has been investigated in the form of unidirectional or bidirectional coupling [8-12] with the goal of developing high-bit-rate secure optical communications over few GHz.

On the other hand, polarization synchronization was reported in unidirectionally coupled vertical-cavity surface-emitting semiconductor lasers (VCSELs). In this case, chaos synchronization occurred when the polarization of the master laser was perpendicular to that of the free-running slave laser [13]. As for mutually coupled VCSELs, in-phase and antiphase synchronization among polarization-resolved chaotic waveforms were demonstrated [14]. Anti-phase chaotic-synchronization dynamics among orthogonally polarized modes was also reported in VCSELs subjected to delayed feedback [15]. Chaos synchronization of orthogonally polarized modes through cross-saturation of population inversions was reported in a solid-state laser [16]. Most recently, chaos synchronization was found to take place for a single pair of transverse modal fields polarized along the critical angles $\theta = \pm\theta_c$ in quasi-locked states of orthogonally polarized transverse modes in a dual-polarization thin-slice solid-state laser with coated end mirrors (abbreviated as TS³L), where the field components of the orthogonally polarized eigenmodes along $\pm\theta_c$ coincide [17].

This Letter reports different forms of lasing patterns, which reflect the relative spatial symmetry of a dimer of coupled orthogonally polarized transverse modes in a TS³L, and verifies the relation between the polarization dependent modal structural change (PSC) on the equator of the Poincaré polarization sphere and the form of chaos synchronization. The unique feature of chaos synchronization among different polarizations is demonstrated in non-reversing mirror symmetric PSC, featuring a distinct difference from the previously reported mirror symmetric PSC [17]. It is shown that there is a high degree of chaos synchronization among transverse mode pairs with parity-symmetric polarizations, which inherits the polarimetric noise properties of the free-running laser. The chaotic dynamics of the whole system consisting of qualitatively different dynamics, which depend on the polarization crossing angle, exhibits similar chaotic dynamics of a single mode laser. A self-organized sender-receiver type of chaos synchronization of a single pair of modes among an infinite number of parity-symmetric polarizations and the total output, featuring their dynamical non-independence, is identified. The associated chaotic camouflaging and extracting of self-mixing solid-state laser Doppler velocimetry signals are demonstrated.

The unique form of polarimetric pattern formation and chaos synchronization as well as chaotic camouflaging self-mixing metrology provides new insights into lasing pattern formation in vector TS³Ls and an application to self-mixing metrologies with polarimetric keys.

The experimental setup is shown in Fig. 1(a). A nearly collimated lasing beam from a laser diode (wavelength: 808 nm) was passed through an anamorphic prism pair to transform an elliptical beam into a circular one, and it was focused on a thin-slice laser crystal by using a microscopic objective lens with a numerical aperture (NA) of 0.5. The laser crystal was a 3-mm-diameter clear-aperture, 1 mm-thick, 3 at%-doped c-cut Nd:GdVO$_4$ whose end surfaces were directly coated with dielectric mirrors $M_1$ (transmission at 808 nm > 95%; reflectance at 1064 nm = 99.8%) and $M_2$ (reflectance at 1064 nm = 99%). Lasing optical spectra were measured by a monochromator (Nikon P250) for obtaining global views and a scanning Fabry-Perot interferometer (SFPI) (Burleigh SA$^{PLUS}$; 2 GHz free spectral range; 6.6 MHz resolution) for measuring detailed structures. The highly sensitive self-mixing modulation was carried out using Doppler-shifted scattered light from a rotating cylinder toward the TS³L cavity [5], as depicted in the figure. The polarization resolved waveform and power spectrum measurement scheme are enclosed by the dashed line.

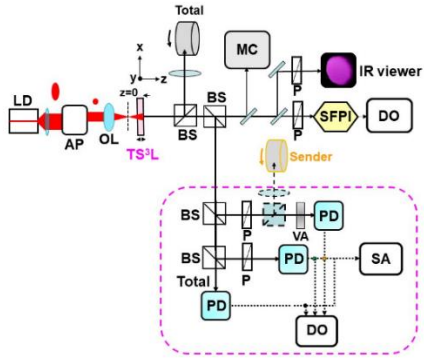

**Fig. 1.** Experimental apparatus. LD: laser diode, AP: anamorphic prism pair, OL: objective lens, BS: beam splitter, VA: variable attenuator, P: polarizer, MC: monochromator, SFPI: scanning Fabry-Perot interferometer, PD: photo-diode, DO: digital oscilloscope, SA: spectrum analyzer (Tektronix 3026, DC–3 GHz).

In the case of TS³Ls, a stable resonator condition is achieved through the thermally induced lensing effect [1], and the input-output characteristics as well as the transverse and longitudinal mode oscillation properties depend directly on the focusing condition (e.g., z-dependent pump-focus spot size and shape) of the pump beam on the crystal due to the mode-matching between the pump and lasing mode profiles [18]. The lasing modes within the pumped region reflected the roughness of the polished surface as well as a tilt of TS³L cavity, where the surface quality of the polished crystal was standard 10/5 scratch/dig per MIL-O-13830, i.e., 1-μm-wide line defects and 50-μm-diameter hole defects [19].

In the experiment, the pump-beam diameter was changed by shifting the laser crystal along the z-axis, as depicted in Fig. 1. When $w_p$ was around 50 μm, dual-polarization oscillations (DPOs) were observed, featuring the Hermite-Gaussian HG$_{00}$ modes as well as their resembling transverse modes. Despite the degenerate HG$_{0,0}$ modes in the cold cavity, thermal birefringence in c-cut vanadate crystals [20, 21] is evident in TS³L, unlike in the case of external cavity lasers whose transverse modes are predominantly determined by the cavity configuration [18]. Moreover, distortion of the transverse mode is caused by the aberration/astigmatism of the thermally induced lens. In fact, when the pump position was precisely changed by moving the laser sample along the x-axis or y-axis in Fig. 1 with an accuracy of 10 μm and a small tilt of the 1-mm-thick cavity of $|a| \leq 1.5°$ was made with an accuracy of 0.3°, as depicted in the inset, DPO featuring deformed HG$_{00}$ modes appeared reproducibly and with different symmetric relations of coupled DPO mode profiles.

Figure 2 show typical examples of PSC on the equator of the normalized Poincaré polarization sphere. $S_{1,2}$ denotes Stokes parameters normalized by the polarization-dependent modal output intensity, $I_p$, i.e., $[S_1, S_2] = [\cos(2\theta), \sin(2\theta)]$ ($\theta$: polarization direction). When pure HG$_{00}$ modes formed for a pair of orthogonally polarized transverse eigenmodes, the polarization rotation is not accompanied by a structural change, as shown in Fig. 2(a) (type I). When a vertically polarized mode at [-1, 0] exhibits a slightly 'elongated' HG$_{00}$ profile, the PSC exhibits mirror symmetry against the $S_1$ axis. Here, the modal profile makes successive structural changes across the vertically polarized elongated HG$_{00}$ mode and returns to the initial mode at [1, 0] against the polarization rotation as depicted in Fig. 2(b) (type II).

By controlling the pump power and position [x, y, z] as well as the tilt of the 1-mm-thick TS³L sample precisely, non-reversing mirror symmetric PSC (type III) shown in Fig. 2(c) was brought about reproducibly. Here, the mirror symmetric PSC like in Fig. 2(b) is prohibited due to the tilt of the elongated HG$_{00}$-like horizontally polarized eigenmode at [1, 0], as depicted by the arrow. Instead, the tilt angle makes a clockwise change toward the vertically polarized eigenmode at [-1, 0] and returns to the initial mode through an anti-clockwise change, as depicted in Fig. 2(c).

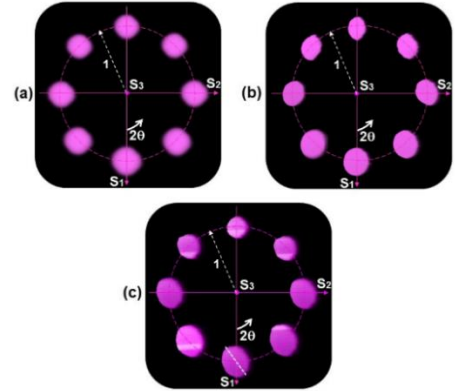

**Fig. 2.** Polarization-dependent modal patterns on the equator of the Poincaré polarization sphere observed with a PbS phototube followed by a TV monitor. (a) Type I, Pump power, P = 151 mW. Threshold pump power, $P_{th}$ = 50 mW. (b)Type II (**Visualization 1**), P = 175 mW, $P_{th}$ = 75 mW. (c) Type III (**Visualization 2**), P = 175 mW, $P_{th}$ = 75 mW.

The global oscillation spectrum measured by the monochromator indicated a single longitudinal mode at λ = 1065.70 nm ($\sigma_2$-transition line; $^4F_{3/2}(2) \rightarrow ^4I_{11/2}(2)$), while the adjacent longitudinal mode separated by $\Delta\lambda = \lambda^2/2nL = 0.258$ nm was not observed (L: crystal thickness, n: refractive index) in all cases.

In type I PSC, arbitrary polarized modal outputs exhibited common behavior even in the chaotic regime because the laser acts as an all-in-one coherent mode through phase locking of orthogonally polarized modes. As for type II PSC, chaos synchronization took place only for a single pair of 'vanishing polarization modes' with $\theta_c \cong \arctan(\pm 1/r)$ (r: amplitude ratio of DPO modes) in the quasi-locked state [17].

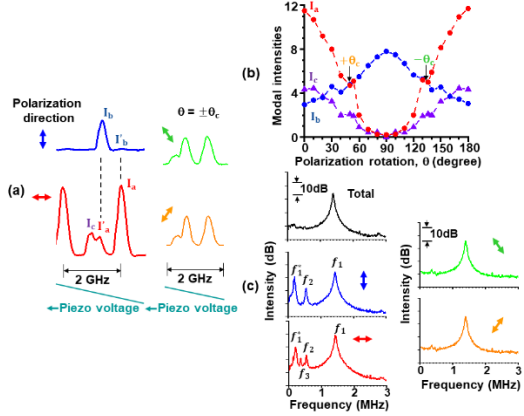

**Fig. 3.** Basic characteristics of type III. P = 160 mW. $P_{th}$ = 67 mW. (a) Optical spectra of DPO modes and vanishing polarization modes. (b) Polarization-dependent modal intensities. (c) Power spectra corresponding to optical spectra.

Now, let me show basic properties of type III PSC below. Figure 3(a) show optical spectra for a pair of DPO eigenmodes (on the left) and for a pair of vanishing polarization modes polarized along $\pm\theta_c$ (on the right). In this case, another mode at the $\lambda$ = 1063.80 nm ($\sigma_1$-transition line, $^4F_{3/2}(2) \rightarrow {}^4I_{11/2}(2)$) joined the $\sigma_2$-transition DPO modes. The $I_c$ mode with a virtual frequency separation of 280 MHz from the $I_b$ mode in the SFPI trace corresponds to the $\sigma_2$-transition mode, assuming a free spectral range of 2 MHz and real oscillation wavelengths of 1063.80 nm and 1065.70 nm. Here, the frequency detuning of 700 MHz remained between the coupled orthogonally polarized eigenmodes. However, it will be shown later that the observed orthogonally polarized transverse modes are not independent and they also form a quasi-locked state.

Figure 3(b) shows the polarization-dependent symmetrical changes of the modal intensities with respect to $\theta$ = 90°. Unlike type II PSC, the polarization-resolved patterns along $\pm\theta$ (i.e., parity-symmetric polarizations) are expected to exhibit synchronized behavior. Indeed, noise-driven relaxation oscillation waveforms for modal components with parity-symmetric polarizations exhibited synchronization.

The power spectra corresponding to Fig. 3(a) are shown in Fig. 3(c), where each power spectrum was obtained by averaging 100 power spectra measured at interval of the update of 160 µs. Here, the strongest noise peak $f_1^*$ below $f_1$ arises due to the $I_a-I_b$ interaction through cross saturation in the transverse direction, while $f_2$ and $f_3$ appear presumably due to the $I_{a(b)}-I_c$ and $I_{a(b)}-I_{a(b)}'$ interactions through cross saturation in the longitudinal direction, where $f_1 = (1/2\pi)[(P/P_{th} - 1)/\tau\tau_p]^{1/2}$ is the relaxation oscillation frequency ($\tau$: fluorescence lifetime, $\tau_p$: photon lifetime). These noise peaks are strongly suppressed both in the total output and a pair of modes polarized along $\pm\theta_c$ due to the combined effect of the self-organized relaxation oscillation dynamics inherent to multimode lasers [22] and the phase-sensitive coherent modal coupling [17]. This suggests that vanishing polarization modes, formed by superposition of orthogonally polarized eigenmode fields with equal amplitude, as well as the whole system behave like a single mode laser, whose noise spectra peak only at $f_1$.

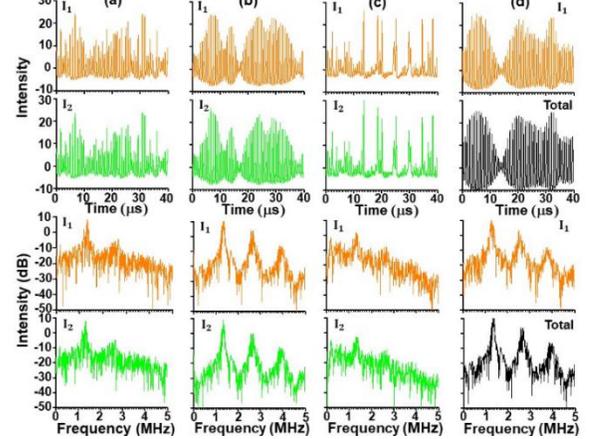

**Fig. 4.** Modal output waveforms and the corresponding power spectra. P = 160 mW. (a) $\theta_{1,2} = \pm 7.5°$, (b) $\theta_{1,2} = \pm\theta_c = \pm 53.5°$, (c) $\theta_{1,2} = \pm 82.5°$, (d) $\theta_{1,2} = 47.5°$, while $I_1$ is synchronized with $I_2$.

Next, a self-mixing TS³L modulation [23] was carried out, where the total output was irradiated onto the rotating cylinder in Fig. 1. Here, simultaneous measurements for modal and total outputs were carried out by using AC-coupled InGaAs photodiodes (New Focus 1812, 30kHz-1GHz) connected to a digital oscilloscope (Tektronics TDS 3052, DC-500 MHz). Chaotic relaxation oscillations were obtained by tuning the Doppler-shift frequency $f_D = 2v/\lambda$ (v: speed along the laser axis) to near $f_1$. The observation time in the whole experiment described below was 100 µs for $10^5$ data. Several examples are shown in Figs. 4(a)-4(d) for pairs of modes with parity-symmetric polarizations along $\theta_{1,2} = \pm\theta$. Qualitatively different chaotic dynamics, which inherited noise properties without modulations, appeared for different $\pm\theta$ values.

It is expected that the total output power spectrum is given by the sum of the power spectra of the vector laser subjected to polarization-resolved optical feedback. Moreover, it is interesting to point out that chaotic waveform of the total output was found to synchronize with those of vanishing polarization modes polarized along $\pm\theta_c$ whose power spectra resemble that of a chaotic single mode laser with spectral components below $f_1$ being eliminated as shown in Fig. 4(b). In return, the chaos synchronization with the total output failed for synchronized pairs of modes with parity-symmetry polarizations except for $\theta_{1,2} = \pm\theta_c$, as shown in Fig. 4(d).

Figures 5(a)-5(d) show correlation plots corresponding to Figs. 4(a)-4(d), where the amplitude correlation coefficient is given by R = $\Sigma_i(I_{1,i} - <I_1>)(I_{2,i} - <I_2>)/\Sigma_i(I_{1,i} - <I_1>)]^{1/2}\Sigma_i(I_{2,i} - <I_2>)]^{1/2}$. Figure 5(e) summarizes the measured R values for several polarization crossing angles for pairs of modes, $\Psi = 2(90 - \theta)$, indicating a high degree of synchronization. Chaos

synchronization among total (T) and vanishing polarizations ($\pm\theta_c$) are depicted by double arrows in the left of Fig. 5(f).

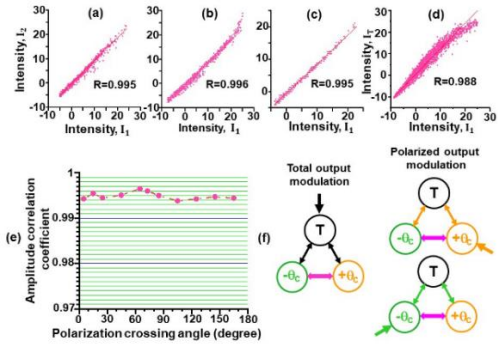

**Fig. 5.** (a)-(d): Intensity correlation plots corresponding to Figs. 4(a)-(d). (e)Amplitude correlation coefficient for different polarization crossing angles. (f) Synchronization diagrams.

Next, the self-mixing modulation dynamics with a polarized output beam impinging on the rotating cylinder was examined as depicted by the dashed inset in Fig. 1. A high degree of chaos synchronization was established only when the power spectra of the pair of modes reproduce well the power spectra of the total output in the free-running condition shown in Fig. 3(c). The vanishing polarization modes polarized along $\pm\theta_c$ satisfy such a requirement. Synchronized chaotic waveforms and the correlation plots are shown in Fig. 6(a), $R_{1,2} \cong R_{2,T} = 0.997$. Figure 6(b) plots the R value versus the polarization crossing angle, $\Psi$. When the laser is perturbed by the $+\theta_c$-polarized mode (namely, sender) subjected to self-mixing modulation, the total output, which synchronizes with the $+\theta_c$-polarized mode, is consider to be excited. Thus, the $-\theta_c$-polarized mode (receiver) synchronized with the total output, whose spectral components below $f_1$ is suppressed similarly to Fig. 4(d), as depicted by double arrows in the right of Fig. 5(f). Consequently, synchronization among the $\pm\theta_c$-polarized modes indicated by the bold double arrow is established.

Note that observed nonlinear dynamics implies the dynamical non-independence of the total and vanishing polarization modal outputs, where synchronization is established among these outputs regardless of the output, which is subjected to the self-mixing modulation, while synchronization with the total output failed for the $\theta \neq \pm\theta_c$ modes.

A polarimetric self-mixing TS³L metrology, which is based on chaos synchronization among sender, receiver and total outputs, is demonstrated in the scheme of Fig. 6(c). A self-mixing laser Doppler velocimetry (LDV) signal from a linearly polarized 0.5-mm-thick a-cut Nd:GdVO$_4$ TS³L was hidden by the chaotic total output of the vector Nd:GdVO$_4$ TS³L discussed above, which was subjected to $+\theta_c$-polarized feedback from the rotating cylinder by inserting a sender's polarimetric "key" (polarizer), as shown in Fig. 6(d). Subtracting the chaotic total output from the camouflaged signal by using the polarized chaotic output in the receiver, the LDV signal at $f_D$ was extracted as shown in Fig. 6(e). Here, the polarized chaotic output synchronized with the total output at $\theta = -\theta_c$ was chosen by a receiver's polarimetric "key" unified with the photodiode.

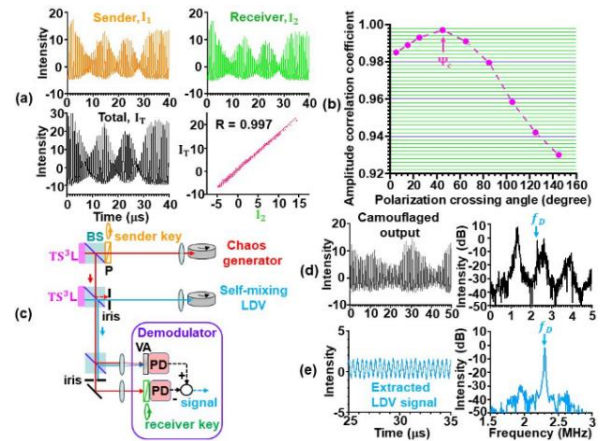

**Fig. 6.** Chaos synchronization induced by self-mixing modulation to one of the vanishing polarization modes: (a) Sender, receiver and total waveforms and correlation plots, (b) Dependence of R on polarization crossing angle $\Psi$. Polarimetric secure TS³L meterology: (c) Experimental setup, (d) Chaotic camouflaging output, (e) Zoom in views of an extracted signal.

**References**

1. L. Fabry, P. Colet and R. Roy, Phys. Rev. A **47**, 4287 (1993).
2. R. Roy and K. S. Thornburg, Jr., Phys. Rev. Lett. **72**, 2009 (1994)
3. G. D. VanWiggeren and R. Roy, Science **279**, 1198 (1998).
4. J. R. Terry, S. Thornburg, Jr., D. J. DeShazer, G. D. VanWiggeren, S. Zhu, P. Ashwin and R. Roy, Phys. Rev. E **59**, 4036 (1999).
5. A. Uchida, M. Shinozuka, T. Ogawa and F. Kannari, Opt. Lett. **13**, 890 (1999).
6. K. Otsuka, R. Kawai, S. L. Hwong, J.-Y. Ko and J.-L. Chern, Phys. Rev. Lett. **84**, 3049 (2000).
7. R. Lang and K. Kobayashi, IEEE J. Quantum Electron. **QE-16**, 347 (1980).
8. H. Fujimoto and J. Ohtsubo, Opt. Lett. **25**, 625 (2000).
9. I. Fischer, Y. Liu and P. Davis, Phys. Rev. A **62**, 011801 (2000).
10. T. Heil, I. Fischer, W. Elsässer, J. Mulet and C. R. Mirasso, Phys. Rev. Lett. **86** 795 (2001).
11. Y. Liu, Y. Takiguchi, P. Davis and T. Aida, Appl. Phys. Lett. **80**, 4306 (2002).
12. J. Ohtsubo, IEEE J. Quantum Electron. **38**, 1141 (2002).
13. Y. Hong, H.-W. Lee, P. S. Spencer and K. A. Shore, Opt. Lett. **29**, 1215 (2004).
14. M. Ozaki, H. Someya, T. Mihara, A. Uchida, S. Yoshimori, K. Panajotov, and Marc Sciamanna, Phys. Rev. E **79**, 026210 (2009).
15. S. Nazhan, "Experimental investigation of anti-phase chaotic-synchronization dynamics of the polarization modes in VCSELs" in Proc. of International Scientific Conference of Engineering Sciences, Published by IEEE, pp.104-107 (2018).
16. K. Otsuka, K. Nemoto, Y. Kamikariya, Y. Miyasaka, J.-Y. Ko, and C.-C. Lin, Phys. Rev. E **76**, 026204 (2007)
17. K. Otsuka, OSA Continuum **2**, 4336 (2019).
18. K. Kubodera and K. Otsuka, J. Appl. Phys. **50**, 653 (1979).
19. K. Otsuka and S.-C. Chu, Laser Phys. Lett. **14**, 075002 (2017).
20. K. Yonezawa, Y. Kozawa and S. Sato, Opt. Lett. **31**, 2151 (2006).
21. J. Lin, V. Petrov, H. Zhang, J. Wang and M. Jiang, Opt. Lett. **31**, 3294 (2006).
22. K. Otsuka, P. Mandel, S. Bielawski, D. Derozier and P. Glorieux, Phys. Rev. A **48**, 1692 (1992).
23. K. Otsuka, Sensors **11**, 2195 (2011).